\documentclass[useAMS,usenatbib,a4]{mn2e}
\usepackage[dv ips]{graphicx}

\title[A new age determination for $\gamma^2$ Velorum]{A new age determination for $\gamma^2$ Velorum from binary stellar evolution models}
\author[John J. Eldridge]{John J. Eldridge$^{1}$ \thanks{E-mail: jje@ast.cam.ac.uk} \\
$^{1}$Institute of Astronomy, The Observatories, University of Cambridge, Madingley Road, Cambridge, CB3 0HA\\}

\pagerange{\pageref{firstpage}--\pageref{lastpage}} \pubyear{2005}
\begin{document}
\maketitle
\label{firstpage}

\begin{abstract}
We derive a new age for the $\gamma^2$ Velorum binary by comparing
recent observations to our set of binary models. We find that it is
very unlikely the stars have not interacted, which implies that
previous estimates of the age from single-star models of $3.5\pm0.4$
Myrs are incorrect. We prefer an older age of $5.5\pm1$ Myrs that
agrees with the age of other lower mass stars in the Vela OB
association. We also find that our favoured binary model shows the
components of the binary have interacted in a Case B, post
main-sequence, mass-transfer event. During the mass-transfer event,
the envelopes of both components where radiative and therefore the
damping of tidal forces are relatively weak. This explains why the
binary is still eccentric after mass-transfer.
\end{abstract}

\begin{keywords}
  binaries: general -- stars: Wolf-Rayet -- binaries: close -- stars:
  individual: $\gamma^2$ Velorum -- stars: fundamental parameters
\end{keywords}

\section{Introduction}

The binary $\gamma^2$ Velorum contains the closest example of a
Wolf-Rayet (WR) star to the Sun. It is a double-lined spectroscopic
binary and an extremely useful tool in understanding the evolution of
Wolf-Rayet stars as the masses, luminosities and mass-loss rate can
all be determined. Recent observations by \citet{north07} have refined
the parameters of the binary orbit and the stellar components. A key
parameter that was estimated was the lower limit age of both stars to
be $3.5\pm0.4$~Myrs.

\citet{jeff09} have took this lower limit of 3.5 Myrs as the age and
suggested that the system was one of the last to form in the Vela OB
association, based on the observation that the other stars are
estimated to be older than 5 Myrs. The age of $\gamma^2$ Velorum was
derived by fitting non-rotating single-star isochrones to the
secondary. This method assumes that the stars have never interacted
and the stars are not rotating. For an average rotation rate the age
would increase by about 25 percent to 4.4 Myrs \citet{mm2000}. Then
assuming no binary interaction initially seems reasonable as the
system has an orbital eccentricity of 0.3 and naively this means that
the system never came into contact. Further consideration however
finds that the orbital separation is small enough, $\approx
100R_{\odot}$, that it is almost impossible for the stars not to
interact through either Roche-Lobe Overflow (RLOF) or the more extreme
Common-Envelope Evolution (CEE). We therefore suggest that the two
stars did interact in a timescale short enough that tidal
circularization either did not occur or was not strong enough to
achieve a completely circular orbit. We note that \citet{north07} do
state that such interaction is possible and implies that their derived
age should only be considered a lower limit. This was not considered
by \citet{jeff09} when they used their age in their study.

We note that the observed parameters of $\gamma^2$ Velorum depend upon
the assumed distance for the system. In this work we have used the
distance derived by \citet{north07} from fitting the binary system of
$336^{+8}_{-7}$~parsecs. This is greater than the Hipparcos distance
of $258^{+41}_{-31}$~parsecs and similar to the distance estimate from
a single measurement by the VLTI that gave a distance to $\gamma^2$
Velorum of $368^{+38}_{-13}$~parsecs \citep{millour}. We have
accounted for the uncertainty in the distance to the binary by
including an error of $\pm0.1$~dex in the luminosities of the binary
components.

There is a large amount of literature showing how binary interactions
can change the evolution of both stars in a binary
\citep{podsibin1,langerbin,vanbmassloss,izzynew,vanbrecent,cantiello,EIT,stancliffe}. However,
while there are many different sources of single-star evolution models
and they mostly agree with each other \citep[e.g.][]{smartt09}, there
are no widely available binary equivalents. Construction of a set of
binary models is not a simple process because rather than only needing
to vary initial mass to obtain different evolution, the initial binary
separation and initial mass of the secondary must also be varied. This
increases the computer time required to cover the full range of
possibly evolutionary paths.

Here we use the set of detailed binary evolution models created by
\citet{EIT} to determine a new age for $\gamma^2$ Velorum. First,
considering the possible ages of the $9M_{\odot}$ WR star. Second,
searching through our binary models for systems that agree with the
masses and orbital separations of the system today. Third, the age of
the secondary star is considered. Afterwards, our results and the
implications of the new age we derive are discussed.

\section{Estimating the age from stellar evolution models}

The stellar models we used were calculated with the Cambridge
STARS code and were described in \citet{EIT}. Here, we restrict ourselves to
use models with a metallicity mass fraction of $Z=0.020$, which we
take to be Solar. These models are all based on circular
orbits. However as discussed by \citet{HPT02} circular and eccentric
orbits with the same semi-latus rectum should produce equivalent
evolution.

\subsection{The age of the Wolf-Rayet star}

It is difficult to accurately determine the age of a WR star. The star
has lost at least half of its initial mass which could have been 20 to
$100M_{\odot}$. However it is possible to estimate an age range for a
WR star from its current mass and WR subtype. The star is designated
as WR11 in the catalogue of \citet{wrcat7} and is identified as a WC
star. This means it is a highly stripped star that has lost all its
hydrogen and most of its helium with the atmosphere dominated by
carbon. We therefore search through our stellar models and record the
age of WC stars with masses between 8.4 and 9.6$M_{\odot}$, the
inferred mass of the WR star in $\gamma^2$ Velorum. Here we consider
the WR star alone and not the companion or binary orbit. We assume a
model is a WC star when there is no hydrogen present and $(X_{\rm
  C}/3+X_{\rm O}/4)/Y \ge 0.03$, where $X_{\rm C}$, $X_{\rm O}$ and
$Y$ are the mass fraction abundance of carbon, oxygen and helium
respectively \citep{mm1994}.

We find that our models indicate a range of possible ages from 3.3 to
6.4~Myrs for WC stars with masses similar to that of the WR star in
$\gamma^2$ Velorum. Our lower estimate of 3.4 Myrs agrees with the age
derived by \cite{north07} for the O star, although this was with older
single star isochrones. The youngest WC stars are from stars with
initial masses above $60M_{\odot}$ and the older WC stars are from
stars with an initial mass of $30M_{\odot}$.  star. Lower initial
masses are preferred by the initial mass function thus initially less
massive, and therefore older, stars are the preferred when we estimate
the age of WR11.

\subsection{The possible initial parameters for the $\gamma^2$ Velorum binary}

\begin{table*}
  \caption{The systems in our grid of binary models that match
    $\gamma$-Velorum. We list their initial parameters and the range
    of the parameters when they match the observed details of
    $\gamma$-Velorum.}
  \label{initialp}
\begin{tabular}{@{}ccccccccccc@{}}
\hline
\hline
Initial          & Initial         &  Initial          & $M_{\rm WR}$  & $M_{\rm O}$   & Age & & \\
$M_{1}/M_{\odot}$ & $M_{2}/M_{\odot}$ & $\log(a/R_{\odot})$ &$M_{1}/M_{\odot}$ & $M_{2}/M_{\odot}$ & /Myrs  & $\log(a/R_{\odot})$ & [C/He]$_{\rm WR}$ & [O/He]$_{\rm WR}$ & $L_{\rm WR}/L_{\odot}$ \\
\hline
35 & 31.5 & 2    & 9.6 -- 8.4 & 30.0 -- 29.8 & 5.51 -- 5.59 & 1.81 -- 1.82  & 0.03 -- 0.25 & 0.001 -- 0.031  & 5.24 -- 5.16  \\
35 & 31.5 & 2.25 & 9.6 -- 8.4 & 29.8 -- 29.6 & 5.51 -- 5.59 & 2.06 -- 2.08  & 0.05 -- 0.30 & 0.003 -- 0.045  & 5.25 -- 5.17  \\
40 & 20   & 2    & 9.6 -- 8.4 & 30.0 -- 30.0 & 5.01 -- 5.13 & 1.89 -- 1.91  & 0.27 -- 0.55 & 0.036 -- 0.153  & 5.26 -- 5.20\\
40 & 28   & 2.25 & 9.6 -- 8.4 & 27.5 -- 27.5 & 5.01 -- 5.14 & 2.07 -- 2.09  & 0.28 -- 0.57 & 0.040 -- 0.168  & 5.26 -- 5.20   \\
\hline
\hline
\end{tabular}
\end{table*}

We have searched through our binary models to find those with a
reasonable match for the $\gamma^2$ Velorum system. For a certain
binary model to match, we require that the following are all true at
some point during the lifetime of the binary.

\begin{itemize}
\item The WR star is of WC type.
\item The mass of the WR star is between 8.4 and 9.6$M_{\odot}$.
\item The luminosity of the WR star is between $\log(L/L_{\odot})=5.1$ and 5.3.
\item The secondary mass is between 27 and 30$M_{\odot}$.
\item The orbital separation is between  $\log(a/R_{\odot})=1.8$ and 2.3.
\item Any interaction occurs after the main sequence (Case B mass-transfer).
\end{itemize}

As already discussed the binary is eccentric and all our models have
circular orbits. We therefore assume that, for our binary systems to
fit, it must have a radius somewhere in the range of separations of
the observed binary. The effect of eccentricity would be for
mass-transfer to begin earlier \citep{church} and thus the initial
separation would be greater for more eccentric systems to obtain the
same evolution as we discuss here.

We choose to only consider systems with mass-transfer after the main
sequence (Case B). This is because when mass-transfer events occur on
the main sequence (Case A) they happen on a nuclear timescale and the
tidal forces would have more time to circularise the orbit. Post
main-sequence mass-transfer occurs on a thermal timescale and
therefore tidal forces have less time to circularise the orbit and the
resulting binary remains eccentric.

We do not attempt to model the secondary here beyond matching its
mass. The secondary does lose mass by stellar winds and mass is
transferred from the primary to the secondary. The accretion rate onto
the secondary is limited to $M_2/ \tau_{\rm thermal}$ where $\tau_{\rm
  thermal}$ is the secondaries thermal timescale. If the mass-transfer
rate is greater than this value the excess mass is lost from the
system. We do not include angular momentum transfer and thermohaline
mixing in this model. This is because they produce complex affects on
the secondary, especially angular momentum transfer which will alter
the amount of rotationally induced mixing. This makes the evolutionary
outcome of the secondary uncertain \citep{cantiello,stancliffe}. We
consider the secondary in more detail in Section 2.3.

We list our models that match $\gamma^{2}$ Velorum in Table
\ref{initialp}. The range of initial primary masses and initial
separations are narrow. The initial separations are similar to the
observed separation. This indicates that the widening of the system
due to stellar-wind mass-loss is countered by the mass-transfer event
tightening the binary.

For each of the systems a binary interaction must have
occurred. Therefore we might simply assume that $\gamma^2$ Velorum
should be a circular orbit rather than eccentric. It is worth asking,
how unusual it is for a Wolf-Rayet binary to be eccentric.  Figure 3
in \citet{wrcat7} shows that in general binaries with periods of 30
days or less tend to be circular. However there are a number of
systems with periods between 30 and 100 days that have eccentricities
in the range of 0.3 to 0.6. This is because binaries with periods
below 30 days experience Case A mass-transfer and therefore their
tidal forces have a long time to circularise the orbit. The binaries
with longer periods experience Case B mass-transfer, that proceeds at
shorter thermal timescales giving less time for tidal forces to
circularise the orbit. These facts suggests that $\gamma^2$ Velorum is
not unusual.

From the above, the assumption that binary mass-transfer events must
always circularise the orbit is not true. We have estimated a
circularization timescale for the models in Table \ref{initialp} and
compared this to the time for the CEE. The key to determine the
timescale of circularization in a binary is the damping mechanism for
the tides created in the star. In general stars with a convective
envelope have shorter timescale than those with radiative envelopes by
around two orders of magnitude \citep{zahn,HPT02}. We find that in our
$35M_{\odot}$ model the star is mostly radiative with a radius of
$90R_{\odot}$ at the time of interaction. The model does have a small
convective shell just outside the helium core between 1 to
$4R_{\odot}$ but it is too deep within the star to effectively damp
tides.

Using the details given in \citet{HPT02} and \citet{demink}, we have
estimated the circularization timescale of the binary to be a few
thousand years, similar to the time of the interaction in our
model. The two timescales are similar in order of magnitude, but
because both are uncertain we conclude it is unlikely that complete
orbit circularization is possible. In comparison we find that for
initially tighter binaries mass-transfer occurs on the main-sequence
via RLOF only over a timescale of $>0.1$ Myrs. Therefore there would
be plenty of time for complete circularization to occur.

One final detail to consider in Table \ref{initialp} are the ratios of
surface carbon, oxygen and helium abundances by number from our
models. \citet{demarco} derived the abundance ratios for $\gamma^2$
Velorum from different model atmosphere codes. They found that [C/He]
took values of 0.15 to 0.06 and [O/He] took values of 0.03 to 0.01
respectively. The binaries in Table \ref{initialp} agree with these
inferred compositions. However looking at the models in detail when we
match the [C/He] ratio we underestimate [O/He] by a factor of 3
indicating the rotation may play some role in the evolution of the
primary star during the WR phase.

\subsection{Modelling the secondary}

Modelling the secondary is not an easy task. If the two stars have
interacted in some way it is difficult to know exactly what effect
this would have had on the secondary
\citep{cantiello,stancliffe}. Here we restrict ourselves to study the
effect of rotation on the secondary alone. We see from Table
\ref{initialp} that there is a wide range of initial masses possible
for the secondary. Some possible progenitors accrete a large amount of
mass to reach that observed today. These stars with lower initial
mass, $M_2 = 20$ and $28M_{\odot}$, are less evolved at the time of
mass-transfer and thus have lower luminosity and/or higher surface
temperature at the time when the primary star matches the
observations.

\begin{figure}
\includegraphics[angle=0, width=84mm]{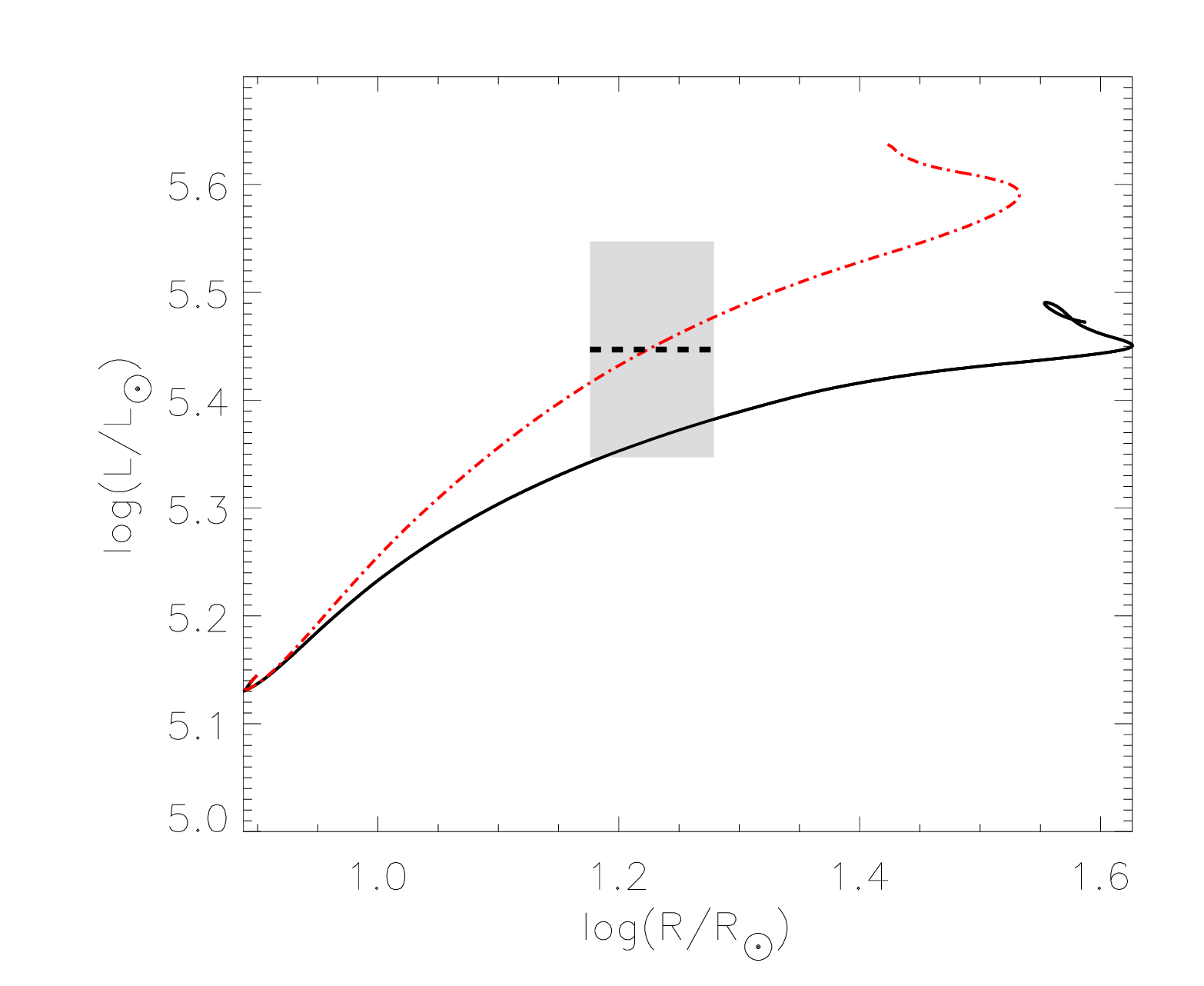}
\caption{The luminosity-radius relationship for two different stellar
  models with initial masses of $31.5M_{\odot}$. The black line
  indicates our normal stellar model while the red dash-dotted line is
  for a stellar model where a small amount of mixing is allowed within
  the radiative zones, simulating the effect of rotation on the
  evolution. The dashed line indicates the parameters of the secondary
  in the $\gamma$-Velorum binary with the grey shaded region
  indicating the error in the luminosity due to the uncertain
  distance.}
\label{figA}

\includegraphics[angle=0, width=84mm]{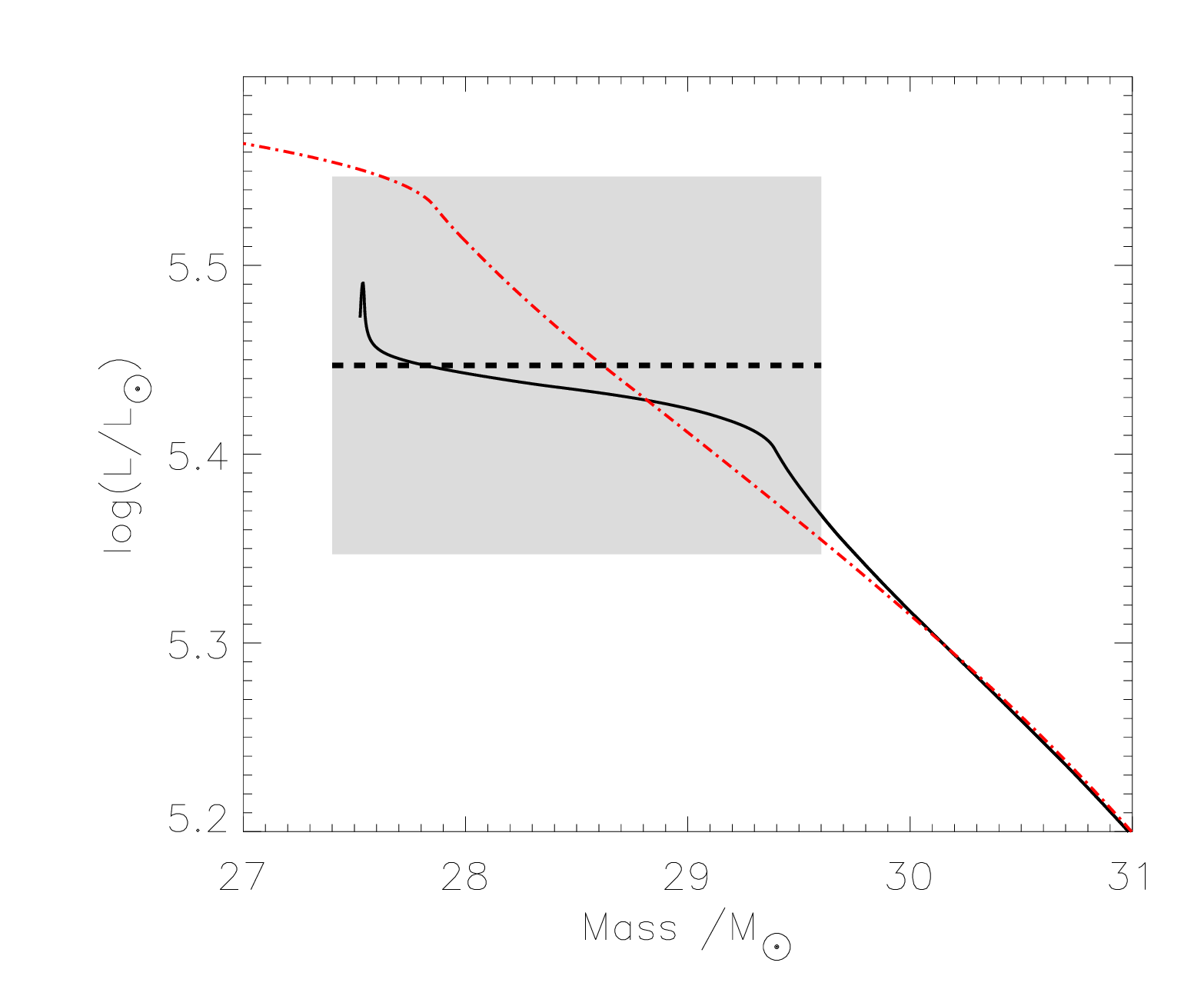}
\caption{Similar to Figure \ref{figA} but the relationship between
  stellar mass and luminosity of our stellar models.}
\label{figB}
\end{figure}

\begin{figure}
\includegraphics[angle=0, width=84mm]{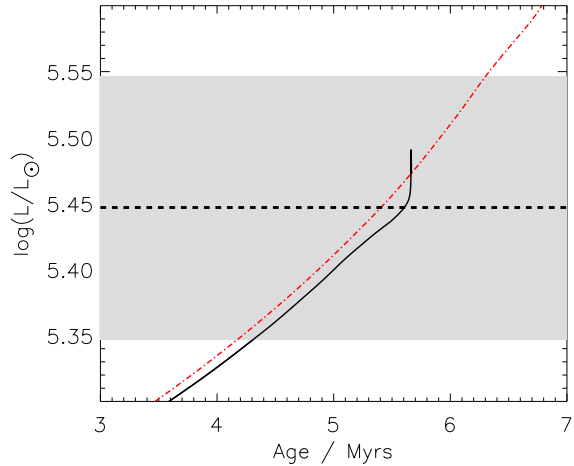}
\caption{Similar to Figure \ref{figA} but the relationship between age
  and luminosity of our stellar models..}
\label{figC}
\end{figure}

Here we concentrate on the secondary star from our favoured model with
an initial mass of $31.5M_{\odot}$ which accretes very little mass. We
find that this star has the same mass as observed for the secondary at
the same time as the primary is a WC star with a mass of $9M_{\odot}$
and fulfills our criterion outlined in Section 2.2. However as shown
in Figure \ref{figA} a basic model only just agrees with the observed luminosity
and radius and $\gamma^2$ Velorum would need to be closer than
300~parsecs. This disagrees with the distances measured by
\citet{north07} and \citet{millour}. If we consider that the star can
only be more distant than we have assumed as indicated by
\citet{millour} and thus more luminous it suggests that our basic
model cannot reproduce the observed secondary radius. In Figure
\ref{figB} we see that the basic model agrees with the luminosity and
mass simultaneously but only at the very end of its main-sequence
lifetime. Using Figure \ref{figC} we estimate an age from
the basic secondary model of $5.6^{+0.1}_{-1.3}$~Myrs.

Rotation is a complicated physical process in stars and our code does
not currently include it. The general effect of rotation is to induce
mixing in a star that increases the time spent on the main-sequence by
mixing nuclear processed material to the surface of the star and new
hydrogen into the core. To simulate this effect we have included an
arbitrary small amount of extra-mixing through the radiative zone of
our model. We see in Figure \ref{figA} that with this mixing the star
achieves a higher luminosity at a given radius and eventually agrees
with the observed luminosity and radius of the secondary. Comparing to
models that do include rotation \citep{mm2005} the evolution is
similar to that provided by reasonable rotation velocities around
$200{\rm km\, s^{-1}}$.

We see in Figure \ref{figB} that the model with enhanced mixing also
agrees with the observed mass and luminosity of the secondary and is
in the middle of its main-sequence evolution. Figure \ref{figC} allows
us to estimate the age of the secondary star to
$5.4^{+0.9}_{-1.2}$~Myrs. This is consistent with the range of ages we
determined for the age of the WC star and binary system above.

\section{Discussion and Conclusions}

We have used three different methods to estimate the ages of
$\gamma^2$-Velorum. These were estimating the age of a $9M_{\odot}$ WC
star, modelling the binary orbit and components simultaneously and
considering the age of the best fitting initial secondary mass with
and without enhanced mixing. We find that the age must be older than
3.5 Myrs and by combining different ages we estimate that it is
$5.5\pm1$ Myrs. The 2 Myrs difference is because \citet{north07} used
single-star non-rotating isochrones to determine a lower limit for the
secondary star age. This work demonstrates the necessity for publicly
available grids of binary star models to be created. The use of
single-star models can give misleading results that lead to
incorrect conclusions.

Our age estimate is in better agreement with ages derived for the
surrounding lower-mass stars from \citet{jeff09}. It suggests that all
the stars in the Vela OB2 association formed at a similar time.

The older age estimate may also help explain the non-detection of
1.8 MeV photon emission from $\gamma^2$ Velorum. This emission is due
to the radioactive decay of $^{26}$Al formed during core hydrogen
burning by proton capture on $^{25}$Mg. $^{26}$Al has a half-life of
0.75~Myrs and $\gamma^2$ Velorum is predicted by \citet{al26} to have
detectable emission. Currently no such emission has been detected. We
suggest there are two factors reducing the amount of $^{26}Al$ below
the detectable abundance. First in our models of $\gamma^2$ Velorum we
find the time difference between the end of core hydrogen burning and
our age estimate is 0.85~Myrs thus less than half of $^{26}Al$ created
remains. Second the lower initial mass of the WR star may mean less
$^{26}$Al is created during the main-sequence as predicted for the
$60M_{\odot}$ star by \citet{al26}.

The fact that the binary is still eccentric does not rule out a binary
mass-transfer event. Our binary models indicate that some post
main-sequence (Case B) mass-transfer events occur so rapidly that the
tides formed in the mostly radiative envelope are unable to
circularise the orbit. On the other hand, WR binaries with periods
below 30 days experiences Case A mass-transfer on the main-sequence
and the tides do have time to circularise the orbit. This places a
limit on the circularization timescale of tides in radiative envelopes
to a few thousand years, the time required for the hydrogen envelope
to be removed in a mass-transfer event.

Our major remaining uncertainty is the evolution of the secondary. We
find that it is possible to explain its current evolutionary state if
initially it was a $31.5M_{\odot}$ star with a rotation rate of
$\approx 200{\rm km \, s^{-1}}$. Less massive stars that accrete a
large amount of mass never match the radius and luminosity of the
secondary when the primary star is a WC star of $9M_{\odot}$. Future
study of this system may provide more clues to the effect of
rotational mixing and of mass-transfer on the secondary stars in
binary systems.

\section{Acknowledgements}
JJE would like to thank the referee Georges Meynet for helpful
comments that lead to a vastly improved paper and the editor for
finding some typos. JJE is currently supported by the IoA's STFC
Theory Rolling Grant. And thanks Rob Jeffries for useful discussion
and Julie Wang for proof-reading.

\bsp



\begin{thebibliography}{99}
\bibitem[\protect\citeauthoryear{Cantiello et al.}{2007}]{cantiello}Cantiello M., Yoon S.-C., Langer N., Livio M., 2007, A\&A, 465L, 29C
\bibitem[\protect\citeauthoryear{Church et al.}{2009}]{church}Church R.P., Dischler J., Davies M.B., Tout C.A., Adams T., Beer M.E., 2009, MNRAS, 395, 1127C
\bibitem[\protect\citeauthoryear{De Marcho et al.}{2000}]{demarco}De Marco O., Schmutz W., Crowther P.A., Hillier D.J., Dessart L., de Koter A., Schweickhardt J., 2000, A\&A, 358, 187D
\bibitem[\protect\citeauthoryear{de Mink et al.}{2009}]{demink}de Mink S. E., Cantiello M., Langer N., Pols O. R., Brott, I., Yoon, S.-Ch., 2009, A\&A, 497, 243
\bibitem[\protect\citeauthoryear{Eldridge, Izzard \& Tout}{2008}]{EIT}Eldridge J.J., Izzard R.G., Tout C.A., 2008, MNRAS, 384, 1109E
\bibitem[\protect\citeauthoryear{Hirschi, Meynet \& Maeder}{2004}]{hmm2004} Hirschi R., Meynet G., Maeder A., 2004, A\&A, 425, 649
\bibitem[\protect\citeauthoryear{Hurley, Tout \& Pols}{2002}]{HPT02}Hurley J.R., Tout C.A., Pols O.R. 2002, MNRAS, 329, 897
\bibitem[\protect\citeauthoryear{Izzard \& Tout}{2005}]{izzynew} Izzard R.G., Dray L.M., Karakas A.I., Lugaro M., Tout C.A., 2006, A\&A, 460, 565I
\bibitem[\protect\citeauthoryear{Jeffries et al.}{2009}]{jeff09} Jeffries R. D., Naylor Tim, Walter F. M., Pozzo M. P., Devey, C. R., 2009, MNRAS, 393, 538J
\bibitem[\protect\citeauthoryear{Maeder \& Meynet}{1994}]{mm1994}Maeder A., Meynet G., 1994, A\&A, 287, 803	
\bibitem[\protect\citeauthoryear{Meynet \& Maeder}{2000}]{mm2000}Meynet G., Maeder A., 2000, A\&A, 361, 101 
\bibitem[\protect\citeauthoryear{Meynet \& Maeder}{2005}]{mm2005}Meynet G., Maeder A., 2005, A\&A, 429, 581 
\bibitem[\protect\citeauthoryear{Millour et al.}{2007}]{millour}Millour F., et al., 2007, A\&A, 464, 107
\bibitem[\protect\citeauthoryear{Mowlavi \& Meynet}{2006}]{al26} Mowlavi N., Meynet G., 2006, NewAR, 50, 484
\bibitem[\protect\citeauthoryear{North et al.}{2007}]{north07}North J. R., Tuthill P. G., Tango W. J., Davis J., 2007, MNRAS, 377, 415N
\bibitem[\protect\citeauthoryear{Podsiadlowski, Joss \& Hsu}{1992}]{podsibin1}Podsiadlowski Ph., Joss P.C., Hsu J.J.L., 1992, ApJ, 391, 246
\bibitem[\protect\citeauthoryear{Smartt el al.}{2009}]{smartt09}Smartt S.J., Eldridge J.J., Crockett R.M., Maund, J.R., 2009, MNRAS in press
\bibitem[\protect\citeauthoryear{Stancliffe \& Eldridge}{2009}]{stancliffe}Stancliffe R.J., Eldridge J.J., 2009, MNRAS submitted
\bibitem[\protect\citeauthoryear{Vanbeveren}{2001}]{vanbmassloss} Vanbeveren D., 2001, in \textit{The influence of binaries on stellar population studies}, Dordrecht: Kluwer Academic Publishers, 2001, xix, 582 p. Astrophysics and space science library (ASSL), Vol. 264. ISBN 0792371046, p.249
\bibitem[\protect\citeauthoryear{Vanbeveren, Van Bever \& Belkus}{2007}]{vanbrecent} Vanbeveren D., Van Bever J., Belkus H., 2007, astro-ph/0703796
\bibitem[\protect\citeauthoryear{van der Hucht}{2001}]{wrcat7}van der Hucht K.A., 2001, NewAR, 45, 135V
\bibitem[\protect\citeauthoryear{Wellstein \& Langer}{1999}]{langerbin}	Wellstein S., Langer N., 1999, A\&A, 350, 148
\bibitem[\protect\citeauthoryear{Zahn}{1977}]{zahn}Zahn J.-P., 1977, A\&A, 57, 383Z

\label{lastpage}
\end{thebibliography}
\end{document}